\documentclass[prl,aps,twocolumn, floatfix, superscriptaddress]{revtex4}
\usepackage{amsmath,bm,graphicx}
\usepackage{amsmath}
\usepackage{amssymb}
\usepackage{amsfonts}
\usepackage{graphicx,color}
\definecolor{r}{rgb}{1,0,0}
\definecolor{b}{rgb}{0,0,1}

\begin{document}

\title{Grid-scale Fluctuations and Forecast Error in Wind Power}

\author{G. Bel}
\affiliation{Department of Solar Energy and Environmental Physics, Blaustein Institutes for Desert Research, Ben-Gurion University of the Negev, Sede Boqer Campus 84990, Israel}
\author{C. P. Connaughton}
\affiliation{Centre for Complexity Science, University of Warwick, Coventry, CV4 7AL, UK}
\author{M. Toots}
\affiliation{Collective Interactions Unit, OIST Graduate University, 1919-1 Tancha, Onna-son, Okinawa, Japan 904-0495}
\author{M. M. Bandi}
\affiliation{Collective Interactions Unit, OIST Graduate University, 1919-1 Tancha, Onna-son, Okinawa, Japan 904-0495}
\email[Corresponding Author: ]{bandi@oist.jp}

\date{\today}

\begin{abstract}
The fluctuations in wind power entering an electrical grid (Irish grid) were analyzed and found to exhibit correlated fluctuations with a self-similar structure, a signature of large-scale correlations in atmospheric turbulence. The statistical structure of temporal correlations for fluctuations in generated and forecast time series was used to quantify two types of forecast error: a timescale error ($e_{\tau}$) that quantifies the deviations between the high frequency components of the forecast and the generated time series, and a scaling error ($e_{\zeta}$) that quantifies the degree to which the models fail to predict temporal correlations in the fluctuations of the generated power. With no {\it a priori} knowledge of the forecast models, we suggest a simple memory kernel that reduces both the timescale error ($e_{\tau}$) and the scaling error ($e_{\zeta}$).
\end{abstract}


\maketitle
\section{Introduction}
Renewable power generation, unlike conventional power, exhibits variability owing to natural fluctuations in the energy source \cite{MacKay2009}. Wind power, in particular, shares spectral features of the turbulent wind from which it derives energy \cite{Apt2007, Milan2013, Calif2014}. This variability in power output adds a cost to renewable power \cite{Lueken2012, Katzenstein2012} that is absent in conventional power sources. Whereas distributed wind farms are expected to smooth the fluctuations \cite{IPCC-Ch7}, power entering the electrical grid still exhibits large amplitude fluctuations \cite{MacKay2009}. Large ramps in power fluctuations present the possibility of grid destabilization \cite{Tande2000} and blackout, a constant source of concern for system operators \cite{Albadi2010, IPCC-Ch7}. This risk increases the cost of operating reserves \cite{Fabbri2005} needed on standby to return a grid back to operation in the event of failure. Naturally, forecast models constitute essential tools in estimating the magnitude of fluctuations beforehand and in planning for the optimal operating reserves required on call. Yet, no standards for forecast accuracy currently exist \cite{Costa2008}.

Extant works on wind power forecast error, ranging from the turbine to the grid scale, focus on modeling the forecast error distribution \cite{Doherty2005, Bludszuweit2008, Hodge2011, Hodge2012a, Hodge2012b, Wu2014}. Since a probability distribution is time-independent, it contains no information on temporal error variations. Several studies have considered the dependence of the mean and the variance of the error on the duration for which the power is predicted (ranging from minutes to hours) \cite{Madsen2009, Lange2005}. Other works have considered the different distributions of errors for mean power over different durations \cite{Hodge2011, Hodge2012a,note1}. However, none of these studies account for the fluctuation correlations of atmospheric turbulence \cite{Katul1994} transferred to the generated power in the analysis of forecast error nor for the temporal correlations of the errors.

Whereas power fluctuations at the scale of an individual turbine \cite{Milan2013} and a wind farm \cite{Apt2007, Calif2014} have been shown to exhibit self-similar scaling, such fluctuations from individual wind farms are expected to smooth out before they enter the electrical grid. Using data from the Irish grid operator EIRGRID \cite{EIGRID}, we show that wind power entering the grid exhibits correlated fluctuations with a self-similar structure. Such scaling points to large-scale correlations in atmospheric turbulence influencing the aggregate wind power entering the grid.

In this article, we exploit these correlations at the grid scale and draw upon the Statistical Theory of Hydrodynamic Turbulence to quantify two types of forecast error. The first is a timescale error ($e_{\tau}$) that quantifies the timescales over which the forecast models fail to predict high-frequency power fluctuations. This timescale error sets a bound on the numerical resolution of forecast models and would already be known to system operators who own and run the forecast models. However, details of the models are not available to potential customers in energy spot markets \cite{Weber2010} who could use this information to factor in the risk associated with a non-supply of promised wind power by producers. The second type of error we quantify is a scaling error ($e_{\zeta}$) that establishes a difference in the self-similar scaling of fluctuations as observed for actual generated power vis {\`a} vis the power that was forecast to be generated. This error could be potentially useful to model developers, and if such an error results from large-scale correlations in atmospheric turbulence, incorporating them into models is not subject to limitations arising from numerical resolution. Having established the errors, we then employ a simple memory kernel upon the forecast time series and show that the errors can be easily reduced with a minimal computational cost.

Two raw time series are provided by EIRGRID: the wind power generated nationwide across all Ireland entering the grid $p_g(t)$, and the power forecast by EIRGRID's models $p_f(t)$ for the same period. The time series sampled at $15$ minute intervals span a five-year period $(2009-2014)$. No other information that permits meaningful data decomposition is available; forecast models employed, the number of wind farms feeding the grid, their location, date of commission or date of scheduled and unscheduled outages, etc. are all unknown. Despite the lack of further information, any quantifiable trends revealed are of immediate use to system operators in estimating operating reserves and in accounting for fluctuations that could potentially destabilize the grid \cite{Paris}. Furthermore, such information \cite{note2} is potentially useful to customers in energy spot markets \cite{Weber2010}.

\begin{figure}
\begin{center}
\includegraphics[width = 2.75 in]{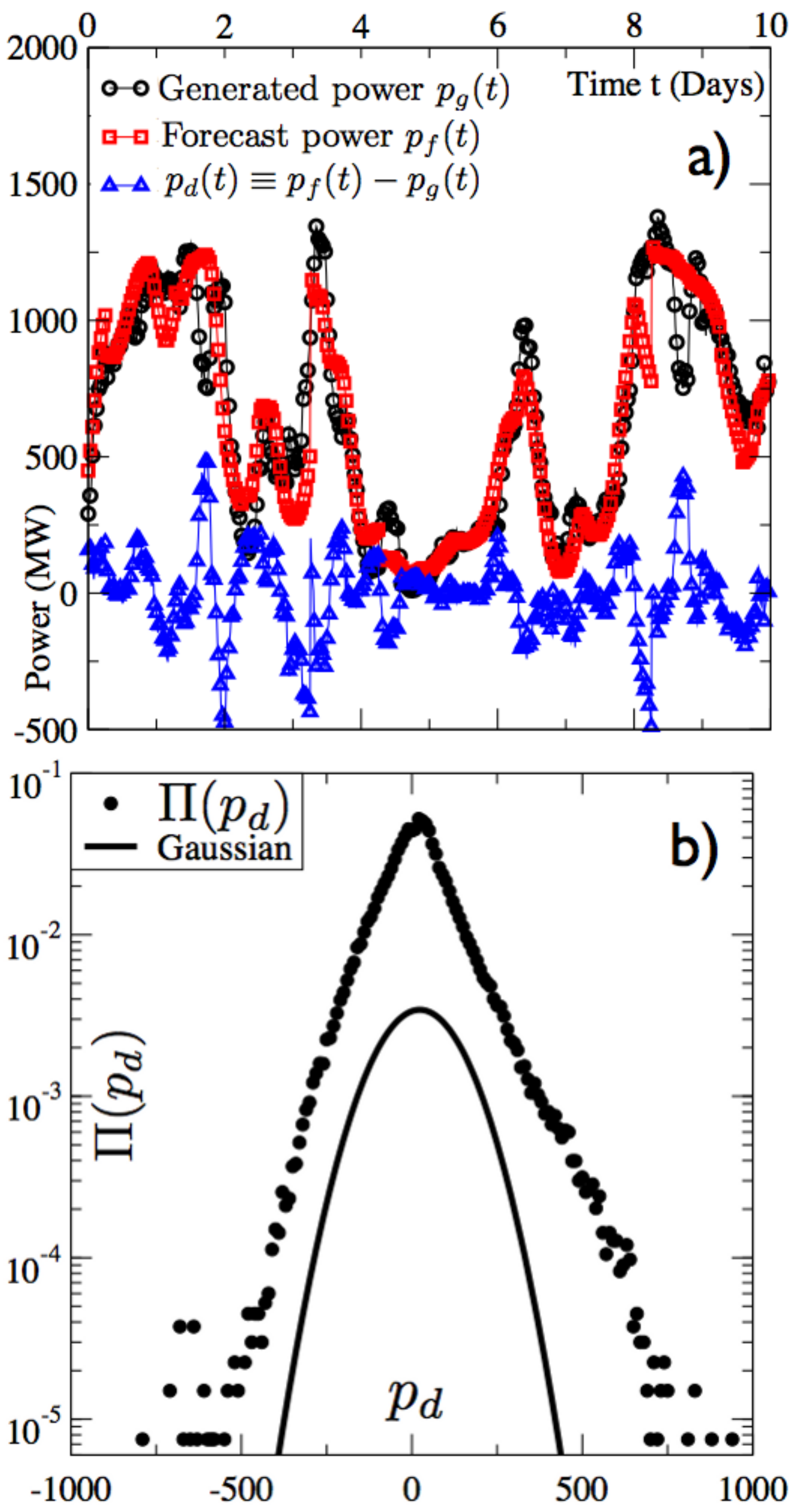}
\end{center}
\caption{(color online) a) Raw time series (for 10 days) of the generated power $p_g(t)$ (black open circles), forecast power $p_f(t)$ (red open squares), and the instantaneous forecast error $p_d(t)$ (blue open triangles) in megawatts (MW). Every third data point is plotted for easy visibility. b) The probability density function of the raw instantaneous forecast error $\Pi(p_d)$ (solid black circles) has exponentially decaying, fat tails relative to a Gaussian distribution (solid black line) of the same mean and standard deviation as $\Pi(p_d)$.}
\label{fig1}
\end{figure}

Raw time series for the generated $p_g(t)$ and forecast power $p_f(t)$, and their instantaneous difference $p_d(t) \equiv p_f(t) - p_g(t)$, which we define as the instantaneous forecast error, are shown in Fig.~\ref{fig1}a for a 10-day period, permitting a few immediate qualitative observations. Firstly, $p_g(t)$ exhibits correlated fluctuations. Secondly, $p_f(t)$, while closely following $p_g(t)$, misses the high frequency (relative to the sampling rate of time series) components. The instantaneous forecast error $p_d(t)$ exhibits correlated fluctuations with a coefficient of variation--standard deviation/mean = $116.24/22.9 \sim 5$, implying large magnitude fluctuations in $p_d(t)$ (i.e., a broad distribution).

\section{Distribution of Forecast Error}
As a point of comparison with prior works \cite{Doherty2005, Bludszuweit2008, Hodge2011, Hodge2012a, Hodge2012b, Wu2014}, we note that the probability density function (PDF) of the raw instantaneous forecast error $\Pi(p_d)$ (fig.~\ref{fig1}b) exhibits fat exponential tails that decay slower than a Gaussian function of the same mean and standard deviation as $\Pi(p_d)$. Indeed, a PDF with exponential tails may be expected for reasons detailed in the following. Being a scalar product of an instantaneous force ($\vec{f}(t)$) and velocity ($\vec{v}(t)$), the statistics of temporal variation in power are determined by the product of two random variables $p(t) \equiv \vec{v}(t) \cdot \vec{f}(t)$ \cite{note3}. The statistics of the product ($Z = XY$) of two normally distributed random variables ($X$ and $Y$) was first studied by C. Craig \cite{Craig1936} (henceforth referred to as Craig's-XY distribution). An asymptotic analysis reveals that Craig's-XY distribution is logarithmically singular about zero with exponentially decaying tails, and its asymmetry (skewness) depends on the instantaneous cross-correlation between $X$ and $Y$ \cite{Bandi2008, Bandi2009}. Whereas this asymptotic analysis is possible only when $X$ and $Y$ are Gaussian, the structure of Craig's XY-distribution itself is more generally observed, even when $X$ and/or $Y$ are non-Gaussian \cite{Bandi2008}.

The basic structure of Craig's-XY distribution is expected for the PDF of power $p(t)$ (e.g.,  compare Fig.~2c in \cite{Milan2013} with Fig.~2 in \cite{Bandi2009}) and its estimation error $\delta p(t)$. Suppose for a single wind turbine, the errors in estimating or forecasting velocity and force are $\delta \vec{v}(t)$ and $\delta \vec{f}(t)$, respectively, then  $p(t) + \delta p(t) \equiv (\vec{v}(t) + \delta \vec{v}(t)) \cdot (\vec{f}(t) + \delta \vec{f}(t))$. Expanding the RHS permits decomposition into power $p(t) = \vec{v}(t) \cdot \vec{f}(t)$ and its error
\begin{equation}
\delta p(t) = \vec{v}(t) \cdot \delta \vec{f}(t) + \delta \vec{v}(t) \cdot \vec{f}(t) + \delta \vec{v}(t) \cdot \delta \vec{f}(t).
\label{error}
\end{equation}
Consequently, $\delta p(t)$ is a random variable whose statistics are determined by the sum of three terms (Eq.~\ref{error}), each being a product of two random variables. One expects $\delta p(t)$ to exhibit features of Craig's-XY distribution irrespective of whether or not  $\vec{v}$, $\delta \vec{v}$, $\vec{f}$, and $\delta \vec{f}$ are Gaussian. In fact, given that the velocity distribution of atmospheric turbulence is known to follow the Weibull distribution \cite{Seguro2000,Monahan2006}, a numerical approach may become necessary.

The power forecast error statistics can be easily scaled from the turbine to the grid scale. If $M$ wind farms feed power to the grid, and the $i$th farm has $N_i$ turbines, the cumulative error in estimating wind power then equals the instantaneous forecast error $p_d(t)$ for the grid, shown in Fig.~\ref{fig1}a, and is given by:
\begin{equation}
p_d(t) = \Sigma_{i=1}^M \Sigma_{j=1}^{N_i} \delta p_j(t).
\label{griderror}
\end{equation}
Summing the power error statistics over all turbines (across all farms) causes an averaging of fluctuations, starting with the most probable ones that occur around zero, thus smoothing the logarithmic singularity. All these generic features are readily observed for $\Pi(p_d)$ in Fig.~\ref{fig1}b. Beyond contributing to the extant literature \cite{Doherty2005, Bludszuweit2008, Hodge2011, Hodge2012a, Hodge2012b, Wu2014}, the structure of $\Pi(p_d)$ provides no useful information for our analysis and will not be discussed further. Understanding temporal variability (fluctuations) and uncertainty (error) requires analysis of the temporal evolution of the distributions, their moments and multipoint temporal correlation functions. We therefore proceed through a statistical analysis of the temporal correlations in the fluctuating time series for generated and forecast power.

\begin{figure}
\begin{center}
\includegraphics[width = 2.75 in]{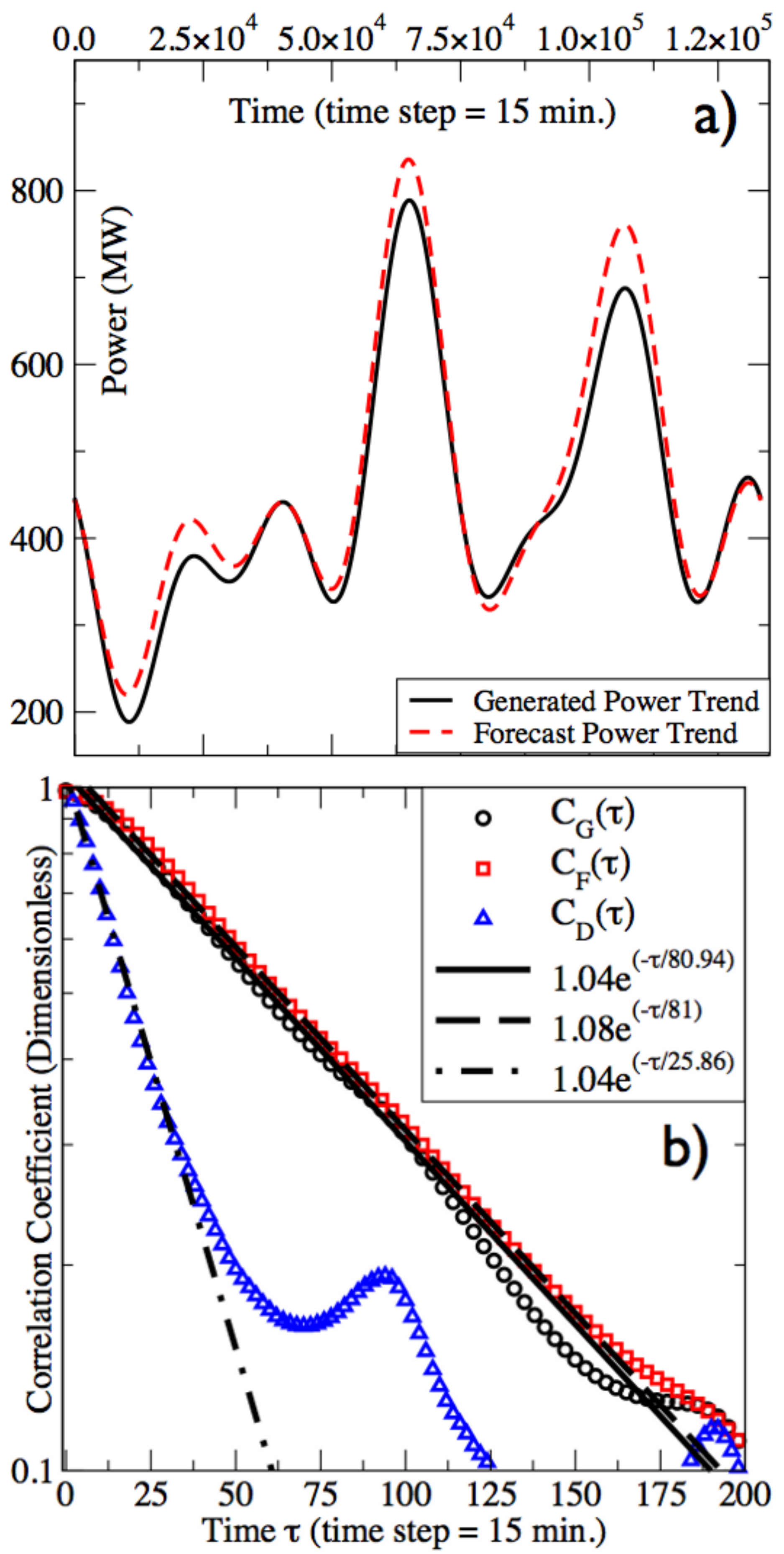}
\end{center}
\caption{(color online) a) The  five-year trend for $p_g(t)$ (black solid line) and $p_f(t)$ (red dashed line) is subtracted from the raw time series in subsequent analysis. b) Log-linear scale: autocorrelation functions $C_G(\tau)$ (open black circles), $C_F(\tau)$ (open red squares) and $C_D(\tau$) (open blue triangles) for $P_G(t)$, $P_F(t)$ and $P_D(t)$, respectively, exhibit exponential decorrelation with respective characteristic timescales obtained from fit to data of $\tau_G = 80.94$ points (20.24 hours), $\tau_F = 81$ points (20.24 hours) and $\tau_D = 25.86$ points ($\sim$ 6.5 hours). Every third data point is plotted for easy visibility.}
\label{fig2}
\end{figure}

\section{Data Analysis}
The time series was analyzed in two stages, with trends in the series being identified in the first stage, followed by an analysis of the fluctuations around the trends in the second stage. Trend removal permits a focus on systematic differences between $p_g(t)$ and $p_f(t)$, ignoring differences due to new wind farms and seasonal variability of the wind power. Trend identification was performed such that the cross-correlation between the generated and forecast power trends was maximal. We used the fast Fourier transform (FFT) for each of the series and defined the trends by inverting the transform using only the frequencies with maximal amplitudes. The number of maximal amplitudes was set by the requirement of the highest cross-correlation between $p_g(t)$ and $p_f(t)$. Keeping the zero frequency (to preserve the signal mean) and five more frequencies resulted in a peak cross-correlation of $0.9904$ between the generated and forecast power trends (Fig.~\ref{fig2}a). These respective trends were subtracted from the raw time series. We denote the de-trended generated power by $P_G(t)$, forecast power by $P_F(t)$ and their instantaneous difference by $P_D(t) \equiv P_F(t) - P_G(t)$.

The characteristic fluctuation timescales for the de-trended time series were first computed from their respective autocorrelation functions defined as:
\begin{equation}
C_X(\tau) = \frac{\overline{ (P_X(t) - \overline{P_X})(P_X(t+\tau) - \overline{P_X})}}{\overline{(P_X(t) - \overline{P_X})^2}}
\label{autocorr}
\end{equation}
where $\overline{P_{X}}$ is a time-average subtracted from the signal (de-trending does not render a zero signal mean since the zero frequency component was preserved). The subscript $X$ should be replaced with $G$ for generated power, $F$ for forecast power, and $D$ for instantaneous forecast error, respectively. The three autocorrelation functions (fig.~\ref{fig2}b) exhibit exponential decay for short times with a data fit following the functional form $C_X (\tau) \sim A_Xe^{-(\tau/\tau_X)}$, where $A_X \simeq 1.0$, owing to $C_X(\tau)$ being normalized, and $\tau_X$ represents the characteristic decorrelation time for each time series, yielding $\tau_G = 80.94$ data points ($\sim 20.24$ hours) for generated power, $\tau_F = 81$ points (also $\sim 20.24$ hours) for forecast power, and $\tau_D = 25.86$ points ($\sim 6.5$ hours) for instantaneous forecast error. The de-trended series were also split into independent time series of shorter duration (1/8th of the original temporal duration). Autocorrelation functions computed for these windowed data did not reveal a measurable difference in the characteristic decay time $\tau_X$; deviations were apparent only for long-term behavior spanning a week (or longer timescales) when the decorrelation had already occurred. The correlation time of high frequency fluctuations ($\lesssim 20$ hours) is much shorter than the slow varying trend (over months to years). Hence the de-trending protocol (in particular, the number of maximal amplitudes) does not influence the analysis to follow--a fact verified and reported upon later.

Autocorrelation functions for the generated ($C_G(\tau)$) and forecast($C_F(\tau)$) power exhibit nearly identical scaling and the same characteristic decay timescales ($\tau_G = \tau_F = 20.24$ hours), suggesting the accurate capture of correlations in generated power by the forecast models. Yet, the autocorrelation function $C_D(\tau)$ for instantaneous forecast error $P_D(t)$ informs us that some correlations are not captured. In particular, we qualitatively know that $P_F(t)$ misses the high frequency components of $P_G(t)$, and they end up in $P_D(t)$, thereby contributing to its two-point correlator. This correlation deficit suggests that higher order moments of the two-point correlator are necessary to capture the statistical structure of the missing fluctuations.

\section{Temporal Structure Functions}
Statistical analysis of higher order correlations is a well-developed, mature tool within the Statistical Theory of Hydrodynamic Turbulence in which higher order two-point correlators are studied through {\it Structure Functions}. Kolmogorov's theory of 1941 (K41) \cite{K41} lays the foundation for structure functions through the celebrated ``4/5th law'': $S_3(r) \equiv \langle (\Delta v_{||}(r))^3 \rangle
 \equiv \langle (v_{||}(R+ r) - v_{||}(R))^3 \rangle = -\frac{4}{5} \overline{\varepsilon} r$, where the third moment of longitudinal velocity differences ($\langle (\Delta v_{||}(r))^3 \rangle$) between two points spatially separated by a longitudinal distance $r$, is proportional to the product of the average turbulent dissipation rate ($\overline{\varepsilon}$) and the longitudinal spacing $r$ \cite{Frisch}. 

The {\it n}th order structure function encodes all cross-terms up to order $n$ of the two-point correlator for a given stationary signal. The physical relevance of structure functions may be appreciated by considering a stationary, fluctuating signal $x(t)$ with zero mean. The difference between the two values of this signal taken time $\tau$ apart ($\Delta x(\tau) \equiv x(t+\tau) - x(t)$) is collected at various windows (of duration $\tau$) along the time series. $\Delta x(\tau)$ is therefore a random variable with statistics of its own, and the {\it n}th order structure function defined as $S_n(\tau) = \langle (\Delta x(\tau))^n \rangle$ is the {\it n}th moment for its PDF $\Pi(\Delta x(\tau))$. The moment $S_n(\tau)$ varies with the time difference $\tau$ between signals, and its scaling (if any) reveals temporal variations in the statistical structure of fluctuations in the signal to the $n$th order.

\begin{figure}
\begin{center}
\includegraphics[width = 2.85 in]{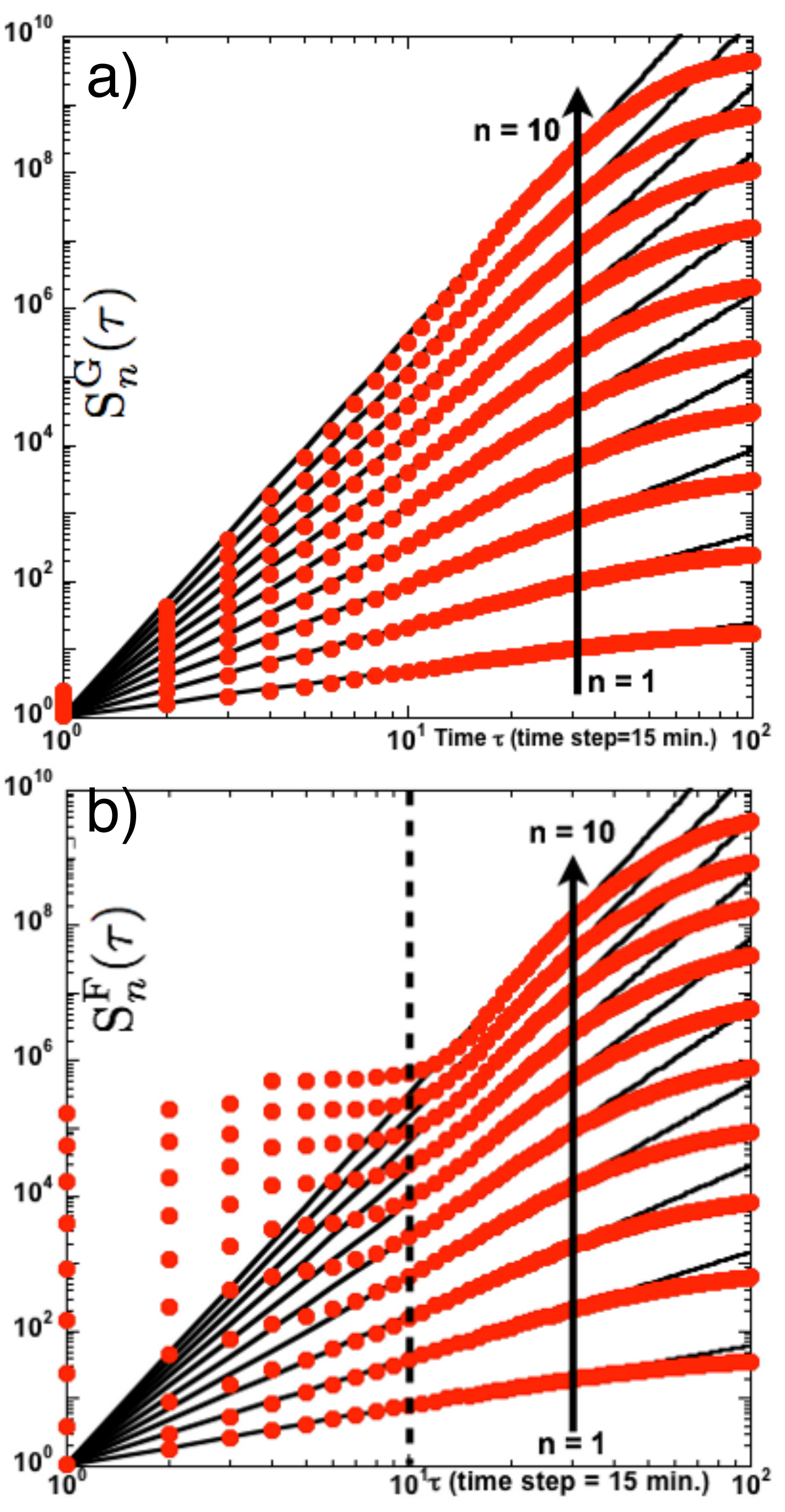}
\end{center}
\caption{(color online) Structure functions of order $n = 1 - 10$ (red solid circles) and their power-law fits (black solid lines) for (a) generated power $S_{n}^G(\tau)$ and (b) forecast power $S_{n}^F(\tau)$ plotted versus $\tau$ in log-log scale exhibit self-similar scaling $S_n^X(\tau) \propto \tau^{\zeta_n^X}$ ($X$ is $G$ for generated and $F$ for forecast power). The scaling is robust for (a) the generated power over 1.4 decades (40 time steps). (b) In contrast, for forecast power, the first- and second-order structure functions exhibit scaling up to $\tau = 40$ time steps, but for $n > 2$, no scaling is observed for $\tau \le 10$ time steps. Self-similar scaling is restored over a limited range of timescales $10 < \tau < 40$.}
\label{fig3}
\end{figure}

Tails of the PDF $\Pi(\Delta x(\tau))$ exert themselves with the increasing order {\it n} of the structure function, thus necessitating more data to resolve higher order structure functions. A weak test for resolving the {\it n}th order structure function involves splitting the time series into smaller windows and testing for identical scaling on the truncated series. However, this test only assures stationarity of the statistics. A strong test for the ability to resolve the {\it n}th order structure function requires that first, the moment's integrand $(\Delta x)^n \Pi(\Delta x) \rightarrow 0$ as $|\Delta x| \rightarrow \infty$ \cite{Bandi2006} (required due to finiteness of data), and second, the PDF $\Pi(\Delta x)$ should decay faster than $1/|\Delta x|^{n+1}$ for $|\Delta x| \rightarrow \infty$ or else the integral $\int (\Delta x)^n \Pi(\Delta x)~\text{d}x$ would diverge for large $|\Delta x|$ \cite{Samorodnitsky1994} (test for existence of a PDF's {\it n}th moment). Whereas the two conditions are not independent, the second condition is theoretical and does not depend upon the available statistics. When conducting data analysis, even when the second condition is satisfied, insufficient data can lead to noise and prevent the integrand $(\Delta x)^n \Pi(\Delta x)$ from satisfactorily converging to zero. The first condition is therefore dependent on finiteness of data. Based on both weak and strong tests, we conclude that the EIRGRID data can resolve structure functions up to order $n = 12$; however, only results up to $n = 10$ are presented.

Since even-order structure functions take only positive values, they converge faster than ones with odd order. To overcome this distinction between odd and even orders, we compute the $n$th order structure function of the absolute value of differences: $S_{n}^X(\tau) \equiv \langle |P_X(t + \tau) - P_X(t)|^n \rangle$ where subtraction of mean $\overline{P_X(t + \tau)}$ and $\overline{P_X(t)}$ is assumed. While ensuring the same convergence rate for even- and odd-order statistics, it also collates all data in the positive quadrant permitting easy visualization. Analysis of fractional-order structure functions allows better testing for anomalous scaling \cite{Chen2005}. Fractional-order structure functions are only defined for absolute values of signal differences \cite{Chen2005}--another reason why we calculate structure functions of absolute differences. We also calculated the structure functions of orders $n = 0.1 - 0.9$ in steps of $0.1$.

\section{Results}
Figure~\ref{fig3} plots the structure functions of order $n = 1 - 10$ (fractional-order structure functions are calculated but not shown) for the absolute value of signal differences of the generated power $|\Delta (P_G(\tau))|$ (fig.~\ref{fig3}a) and forecast power $|\Delta (P_F(\tau))|$ (Fig. \ref{fig3}b). Self-similar or power-law scaling is observed for the generated power structure functions over $1.4$ decades spanning $\tau \le 40$. Scaling over the same temporal range is also observed for the forecast power structure functions of order $n = 1$ and $2$. For $n > 2$, no scaling is observed for timescales $\tau \le 10$. The scaling is restored over a limited range of timescales $10 < \tau < 40$ ($0.4$ decades in time).

\begin{figure}
\begin{center}
\includegraphics[width = 2.75 in]{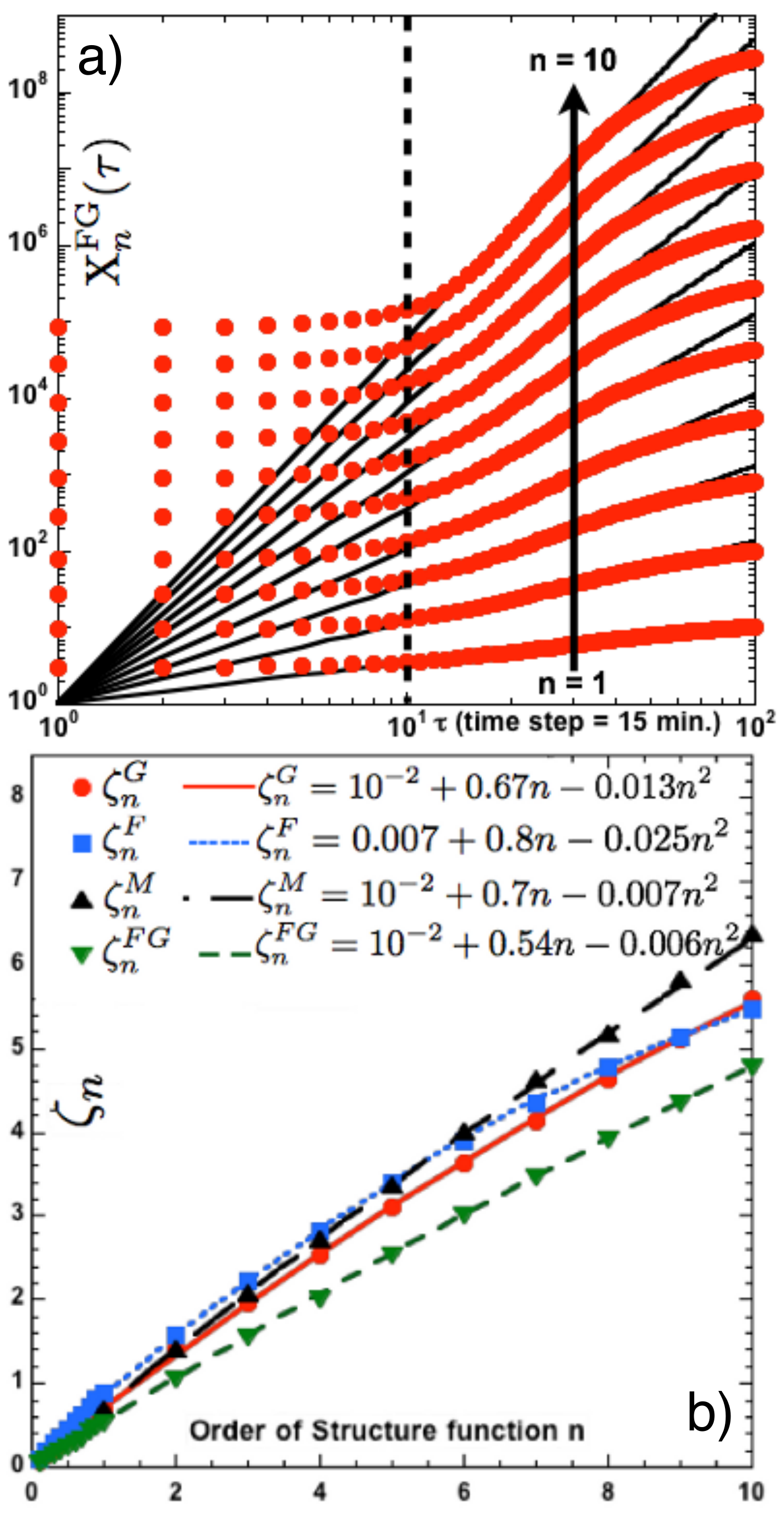}
\end{center}
\caption{(color online) a) Log-log scale: cross-structure functions $X_{n}^{FG}(\tau)$ versus $\tau$ (red solid circles) exhibit no scaling at early times $\tau \le 10$, with scaling restored for $10 < \tau < 40$. Black solid lines are power-law fits to data within the scaling regime. b) Scaling exponent $\zeta_{n}^X$ versus the order of structure function {\it n} for generated $G$ (red solid circles), forecast $F$ (blue solid squares) and modified forecast $M$ (black solid triangles) structure functions, and cross-structure functions $FG$ (green solid inverted triangles), and their respective second-order polynomial fits: solid red line for $\zeta_{n}^G$, small dashed blue line for $\zeta_{n}^F$, medium dashed black line for $\zeta_{n}^M$ and long dashed green line for $\zeta_{n}^{FG}$.}
\label{fig4}
\end{figure}

Self-similar scaling of the temporal structure functions implies a relationship of the form:
\begin{equation}
S_{n}^X(\tau) \propto A_{n}^X \tau^{\zeta_{n}^X}
\label{scaling}
\end{equation}
where $\zeta_{n}^X$ is the scaling exponent. For simple mono-fractal scaling, $\zeta_{n}^X \propto n$. However, fluctuations with a multi-fractal character exhibit a nonlinear dependence of the scaling exponent $\zeta_{n}^X$ with respect to $n$. Super- (sub-) linear variation of $\zeta_{n}^X$ versus $n$ implies temporal expansion (compression) of fluctuations \cite{Mandelbrot2004}. Scaling exponents for all the structure functions were computed from the log derivative, $\zeta_{n}^X = \frac{d~\text{log}(S_{n}^X(\tau))}{d~\text{log}(\tau)}$, which provides a more reliable estimate of the exponent than a power-law fit \cite{Chen2005, Larkin2009}. The pre-factor $A_{n}^X$ in Eq. \ref{scaling} is subsequently obtained from fit to data. In Fig.~\ref{fig3}, all the data (red solid circles) were divided by $A_{n}^X$ such that all fits (solid black lines) commence from both mantissa ($\tau$) and ordinate ($S_n^X(\tau)$) at unity, for an easy comparison of $\zeta_{n}^X$ with order $n$. All the data in Fig.~\ref{fig3}, Fig.~\ref{fig4}a, and Fig.~\ref{fig5}b therefore follow the scaling relation: $S_{n}^X(\tau) \propto \tau^{\zeta_{n}^X}$ ($A_{n}^X \equiv 1$).

The scaling in Fig.~\ref{fig3} reveals higher-order temporal correlations at work in the EIRGRID data. The absence of scaling for $S_{n}^F(\tau)$ for $n > 2$ at timescales $\tau \le 10$ confirms the qualitative observation made in Fig.~\ref{fig1}a that forecast models do not capture high frequency fluctuations. More importantly, Fig.~\ref{fig3}b ascribes a precise bound on the time ($\tau = 10$, 2.5 hours) out to which the high frequency fluctuations are missed. Finally, scaling presence for $S_{n}^F(\tau),~n =1, 2$ explains the close agreement between the autocorrelation functions $C_G(\tau)$ and $C_F(\tau)$ and their identical characteristic decay times, $\tau_G$ and $\tau_F$, observed in Fig.~\ref{fig2}b. This is to be expected on the grounds that the second-order structure function $S_2(\tau) \equiv \langle (\Delta x(\tau))^2 \rangle  = \langle x(t+\tau)^2 \rangle + \langle x(t)^2 \rangle - 2 \langle x(t)x(t+\tau)\rangle$ shares a direct correspondence with the autocorrelation function where the cross-term is identical to the numerator of Eq. \ref{autocorr}. The failure of $S_{n}^F(\tau)$ for $n > 2$ to capture high frequency fluctuations out to $\tau = 10$ reveals one type of forecast error in the models; we call this the {\it timescale error} $e_{\tau}$.

Before proceeding to the second type of error arising from scaling mismatch, we define the cross-structure function $X_{n}^{FG}(\tau) \equiv \langle |P_F(t+\tau) - P_G(t)|^n \rangle$. $X_{n}^{FG}(\tau)$ represents {\it n}th order moments for the PDF of the relative magnitude of fluctuations between $P_G(t)$ and $P_F(t+\tau)$, and their cross-terms correspond to higher-order two-point cross-correlators between the generated and forecast power. This function is plotted in Fig.~\ref{fig4}a. Again, we notice that scaling is absent at early times ($\tau \le 10$), and restored at later times ($10 < \tau < 40$). We note that  $X_{n}^{FG}(\tau)$ exhibits no scaling for $n = 1$ and $2$, unlike the forecast structure functions (Fig.~\ref{fig3}b). Although $S_{n}^F(\tau)$ exhibits scaling for order $n =$ 1 and $2$, its exponent $\zeta_{n}^F \neq \zeta_{n}^G$; this scaling deficit is reflected in $X_{n}^{FG}(\tau)$ for $n =1$ and $2$.

\section{Discussion}
Having established the various structure functions, we now consider the behavior of their scaling exponents $\zeta_{n}^X$ ($X \equiv G$ for generated, $F$ for forecast and $FG$ for the cross-structure function). Figure \ref{fig4}b plots $\zeta_{n}^X$ versus the order $n$ together with their polynomial fits to the quadratic order. $\zeta_{n}^G = 10^{-2} + 0.67n - 0.013n^2$ scales almost linearly (mono-fractal) with a small, but measurable, quadratic deviation towards multi-fractal behavior. The exponent $\zeta_{n}^F = 0.007 + 0.8n - 0.025n^2$ exhibits a slightly more pronounced quadratic deviation (multi fractal behavior) relative to $\zeta_{n}^G$. On the other hand,  $\zeta_{n}^{FG} = 10^{-2} + 0.54n - 0.006n^2$ scales almost linearly with {\it n}, implying mono-fractal scaling.

We now consider the measurement error for the aforementioned scalings. Firstly, given that all de-trending protocols suffer from an {\it ad hoc} choice of a de-trending timescale, we tested the scalings for dependence on the de-trending procedure by varying the number of maximal amplitudes. Ignoring the condition for maximal cross-correlation between $p_g(t)$ and $p_f(t)$, the number of maximal amplitudes contributing to the trends was varied. The scalings were invariant up to the inclusion of 15 maximal amplitudes into the trend, beyond which, coefficients for the polynomial fits started varying in the second decimal place. Having ascertained the robustness of our choice for the five maximal amplitudes at which the cross-correlation peaks, we focused on a second source of scaling measurement error, namely statistical variability. Since the scalings are analyzed up to $\tau = 100$ data points, the de-trended time series were split into eight independent windows (each with $21912$ data points), and the structure functions were re-computed for each window. The variation in the log derivative ($\zeta_{n}^X = \frac{d~\text{log}(S_{n}^X(\tau))}{d~\text{log}(\tau)}$) for the eight independent measurements was taken as the possible scatter in the scaling estimation, thereby providing a confidence interval for the polynomial fits. The scatter was found to be $\zeta_{n}^X \pm 0.01$ in both the measured value of $\zeta_{n}^X$ and the corresponding polynomial fits (for each of the polynomial coefficients) for each of the eight independent datasets, revealing that the polynomial fits were meaningful only to the linear order for $\zeta_{n}^G$ and $\zeta_{n}^{FG}$. The quadratic-order polynomial coefficient for $\zeta_{n}^F$, despite being larger than the scatter of $\pm 0.01$, is not useful owing to the fact that the corresponding quadratic terms for $\zeta_{n}^G$ and $\zeta_{n}^{FG}$ are smaller than the scatter magnitude.

Despite qualitatively observing a quadratic deviation for $\zeta_{n}^X$ in Fig.~\ref{fig4}b, our inability to ascribe significance to it arises from the fact that the multi-fractal component (deviation from linear scaling) of the scalings is miniscule. This is significant in light of several studies that have demonstrated multi-fractal scaling for wind power fluctuations at the turbine \cite{Milan2013, Calif2014} and farm scales \cite{Calif2013}. Turbulence theory traces the source of multi-fractal behavior to intermittent fluctuations that can arise from two sources in the atmospheric context. The first, known as internal intermittency, occurs at the small scales of turbulent flow. These intermittent fluctuations would be naturally reflected in the power generated at the turbine and farm scales. However, when adding together power generated by geographically distant wind farms, internal intermittency should smooth out \cite{Katzenstein2010} since it is a small-scale effect and cannot extend across geographically distributed wind farms. Furthermore, the sampling interval (15 minutes) for EIRGRID data is not expected to resolve any effects that may arise from internal intermittency, which occur at much shorter timescales (high frequencies).

The second source of intermittency, known as external intermittency, occurs at the edge of any free-stream \cite{Kuznetsov1992} and arises in the atmospheric context due to coupling between the atmospheric boundary layer turbulence and a co-moving weather system \cite{Katul1994}. External intermittency, which can be experienced in the form of wind gusts, is of greater relevance in the present analysis as it can both correlate distributed farms through the weather system and occur at timescales longer than the 15-minute sampling interval for EIRGRID data. The nearly fractal scaling of $\zeta_{n}^G$ informs us that both internal and external intermittency are being smoothed to the point of rendering grid-level power fluctuations almost mono-fractal.

The self-similar scaling of $S_n^G(\tau)$ over several hours does strongly point to the influence of large-scale turbulent structures on power fluctuations at the grid level. The 20-hour characteristic decorrelation time ($\tau_G$) for generated power in Fig.~\ref{fig2}c, if taken as the large eddy turnover time of atmospheric turbulence, also lends credence to such an argument. Finally, independent proof in support of this argument also comes from Katzenstein et al. \cite{Katzenstein2010} who show that an individual wind farm exhibits $f^{-5/3}$ ($f$ being the frequency) scaling for the wind power spectrum (equivalent to $\tau^{2/3}$ scaling of the second-order structure function in the time domain). However, as wind power from various farms is summed, the spectrum steepens (please see Fig. 3 in \cite{Katzenstein2010}). Such spectral steepening can be clearly attributed to the smoothing of high frequency (short timescale) fluctuations corresponding to small eddies. But the low frequency (long timescale) fluctuations corresponding to large-scale eddies lose no power spectral density, clearly indicating the influence of large-scale turbulent structures.

We finally consider the forecast error due to the scaling mismatch. We define the scaling error as $e_{\zeta} \equiv \zeta_{n}^F - \zeta_{n}^G$. Under this definition, if the time series for forecast and generated power were identical, then $S_{n}^G(\tau) \equiv S_{n}^F(\tau)$, implying $\zeta_{n}^G \equiv \zeta_{n}^F$, and therefore $e_{\zeta} = 0$. Another typical case arises if forecast models fail completely, resulting in a flat time series with no fluctuations, $\zeta_{n}^F = 0$ resulting in an error $e_{\zeta} = - \zeta_{n}^G$. Using the polynomial fits for $\zeta_{n}^X$ (see Fig.~\ref{fig4}b) to linear order, we obtain $e_{\zeta} = (7 \times 10^{-2} + 0.8n) - (10^{-2} + 0.67n) = -0.003 + 0.13n$. This can be cross-validated against the difference $\zeta_{n}^G - \zeta_{n}^{FG} = (10^{-2} + 0.67n) - (10^{-2} + 0.54n) = 0.13n $. Since $\zeta_{n}^X \rightarrow 0$ as $n \rightarrow 0$, the 0th order term falling within the scatter may be taken to be zero. Both estimates of error are identical in linear order ($e_{\zeta} = 0.013n$).

\begin{figure}
\begin{center}
\includegraphics[width = 2.85 in]{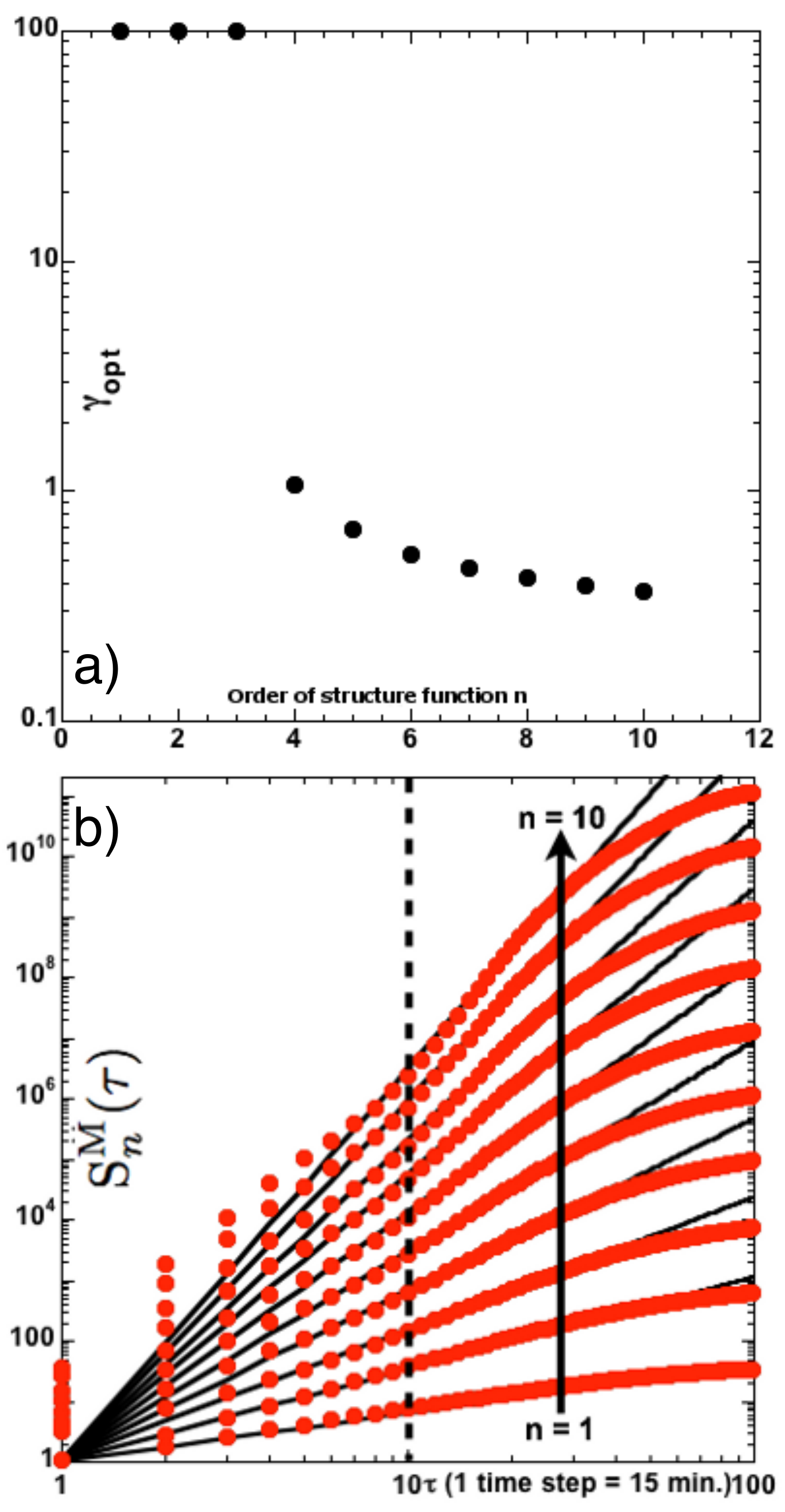}
\end{center}
\caption{(color online) a) Log-linear scale: $\gamma_{\text{opt}}$ versus order of structure function shows no improvement for $n < 4$ but shows better agreement for $n \ge 4$ with an abrupt change observed in $\gamma_{\text{opt}}$ at $n = 4$. b) Log-log scale: Structure functions $S_n^M(\tau)$ versus $\tau$ (red solid circles) for the modified forecast time series show considerable improvement over their counterparts $S_n^F(\tau)$ in Fig.~\ref{fig3}b.}
\label{fig5}
\end{figure}

The analysis thus far demonstrates the importance of temporal correlations in wind power and their role in estimating forecast errors. It is reasonable to ask whether this knowledge could help improve the forecast time series, despite having no knowledge of the models employed. In particular, to capture the short-term correlations missed by the forecast, we introduce a modified forecast that is based on the original forecast, convoluted with an exponentially decaying memory kernel derived from the generated power time series. The modified forecast power is given by $P_M(t)=\int_0^t P_F(\tau)e^{-\gamma(t-\tau)}d\tau$.

The memory duration ($1/\gamma$) was chosen so as to minimize the relative difference between the structure functions of the generated and forecast power. As expected (as shown earlier, the low-order structure functions of the generated and forecast power are very similar), we found that the optimal $\gamma$ varies with the order of the structure function. For $n<4$, the memory-modified forecast shows no improvement in the agreement between $S_{n}^G$ and $S_{n}^F$. For $n\geq 4$, the modified forecast exhibits better agreement with the structure functions of the generated power as shown in Fig.~\ref{fig5}b. The optimal $\gamma$ ($\gamma_{\text{opt}}$) was found to be $\gamma_4\approx 1.06 $ and $\gamma_{10}\approx 0.37$, as shown in Fig.~\ref{fig5}a, plotted in log-linear scale to show the variation in $\gamma_{\text{opt}}$ for $n \ge 4$. The simple scheme, suggested here, not only tries to rectify the timescale error $e_{\tau}$, but also attempts to statistically align the temporal correlations by improving the scaling error $e_{\zeta}$.

As is apparent from Fig.~\ref{fig5}b, the structure functions ($S_n^M(\tau) \equiv \langle |\Delta P_M(\tau)|^n \rangle$) for modified forecast time series are substantially improved over their unmodified counterpart (fig.~\ref{fig3}b). Firstly, scalings are restored at high frequencies ($\tau \le 10$), thus rendering the timescale error irrelevant. More importantly, the scaling itself is improved as is evident from Fig.~\ref{fig4}b, revealing $\zeta_n^M = 0.01 + 0.7n - 0.007n^2$. To linear order, the scaling error $e_\zeta = \zeta_n^M - \zeta_n^G = 0.7n - 0.67n = 0.03n$, a considerable improvement over the original forecast time series. Being computationally inexpensive, and given that spinning and non-spinning reserves must act within 10 minutes of failure, with replacement reserves acting within 20-60 minutes, there are tangible benefits to incorporating such a memory kernel into models to monitor instabilities in real-time. Furthermore, it might be possible to improve the forecast models using different parameterizations of the regional climate models or weather models.

\section{Summary}
In summary, wind power exhibits significant temporal correlations even at the grid level, where fluctuations are expected to average out \cite{IPCC-Ch7} as power is fed from geographically distributed wind farms. Previous studies show that the temporal correlations of the wind are essential to studying wind-generated large-scale ocean currents \cite{Bel2013}; a similar appreciation of large-scale correlations in atmospheric turbulence within the context of wind power is called for. Fluctuations, albeit posing a problem to system operators, possess a statistical structure through temporal correlations, which could be exploited to quantitatively analyze the error in forecast models. The technique proposed here is only limited by the sampling rate of the time series. Beyond potentially serving as a standard for quantifying wind-power forecast accuracy, it could have applications for any renewable energy source with temporally correlated fluctuations possessing a statistical structure.

\acknowledgments
MT and MMB were supported by the OIST Graduate University with subsidy funding from the Cabinet Office, Government of Japan. CPC was hosted by OIST Graduate University while performing this work. GB was supported through the European Union Seventh Framework Programme (FP7/2007-2013) under grant number [293825]. The authors gratefully acknowledge EIRGRID for permission to use their data and N. Ouellette for scientific discussions.

\bibliography{all2}

\begin{thebibliography}{43}
\bibitem[{not({\natexlab{a}})}]{note1}
\bibinfo{note}{Note that \cite{Hodge2011} suggests the Cauchy distribution for
  the errors. However, this distribution is not suitable because all its
  moments are undefined \cite{Levy}}.
\bibitem[{not({\natexlab{b}})}]{note2}
\bibinfo{note}{This does not apply for Ireland since its grid is isolated from
  mainland Europe}.
\bibitem[{not({\natexlab{c}})}]{note3}
\bibinfo{note}{One can define power as $p(t) \equiv A\vec{v}(t)^3$ (where $A$
  is a suitable constant) and apply the same arguments without loss of
  generality}.
\end{thebibliography}

\end{document}